\documentclass[showpacs,twocolumn,prb]{revtex4}
\usepackage{amsmath}
\usepackage{graphicx}
\usepackage{dcolumn}
\usepackage{color}
\usepackage{bm}
\usepackage{bbm}
 \newcommand{\ba}{\begin{eqnarray}}
 \newcommand{\ea}{\end{eqnarray}}
\begin{document}

\title{Spontaneous $\mathcal{PT}$ symmetry breaking and quantum phase transitions in dimerized spin chains}

\author{Gian Luca Giorgi} \email{gianluca@ifisc.uib-csic.es}
\affiliation{Instituto de F\'{i}sica Interdisciplinar y Sistemas Complejos IFISC (CSIC-UIB),
Campus Universitat Illes Balears, E-07122 Palma de Mallorca, Spain}


\begin{abstract}
The occurrence of parity-time reversal ($\mathcal{PT}$) symmetry breaking is discussed in a non-Hermitian spin chain. 
The Hermiticity of the model is broken by the presence of an alternating, imaginary, transverse magnetic field. 
A full real spectrum, which occurs if and only if all the eigenvectors are $\mathcal{PT}$ symmetric, 
can appear only in presence of dimerization, i.e. only if the hopping amplitudes between nearest-neighbor 
spins assume alternate values along the chain. In order to make a connection between such system and the Hermitian
world, we study the critical magnetic properties of the model and look for the conditions that would allow to observe 
the same phase diagram in the absence of the imaginary field. Such procedure amounts to renormalizing the spin-spin coupling amplitudes.
\end{abstract}

\maketitle

Since the paper of Bender and Boettcher,\cite{bender} it is known that non-Hermitian Hamiltonians
can display a real spectrum if they are invariant under the joint action of parity ($\mathcal{P}$) and time reversal ($\mathcal{T}$) symmetry.
The parity operator $\mathcal{P}$ performs spatial reflection and its action consists of changing the sign of both position and momentum, whereas the anti-linear
time-reversal operator $\mathcal{T}$ maps $p$ in $-p$ and the imaginary part $i$ in $-i$.
Non-Hermitian Hamiltonians that are $\mathcal{PT}$ symmetric generate a complex extension to the Hermitian quantum mechanics.\cite{mosta1,mosta2}
The axiom that forces the Hamiltonian $H$ to be Hermitian is in fact introduced in order
to guarantee the existence of a stable ground state and unitary evolutions. However, these two basic requirements 
are still satisfied in the presence of $\mathcal{PT}$-symmetry.\cite{benderrev}

On the other hand, because of the anti-linear character of the parity-time reversal operator, the condition $[H,\mathcal{PT}]=0$
is not sufficient to have a $\mathcal{PT}$-symmetric system. It is in fact possible to observe 
spontaneous symmetry breaking in some of the eigenstates of $H$, associated to the appearance of complex
conjugate eigenvalues. The prototypical model introduced in the literature\cite{bender} is
that of system obeying the Hamiltonian $p^2-(ix)^N$. Its spectrum turns out to be real only if $N\ge 2$. 
As $N$ decreases, the number of complex eigenvalues increases, and for $N\rightarrow1^+$, no real roots are detected.
For $N<2$, the $\mathcal{PT}$ symmetry is spontaneously broken.

The search for non-Hermitian quantum models in discrete systems has led to study tight-binding particle models \cite{jin,jin2}
and spin chains.\cite{korff,kw,castro}

The understanding of such systems on physical bases seems to be crucial, given that the meaning of non-Hermitian Hamiltonians is rather unclear.
Mostafazadeh showed that a necessary and sufficient condition for the reality of all the eigenvalues
is the existence of an ``equivalent Hermitian counterpart'',\cite{mosta3} that can be built starting from the eigenfunctions of the Hermitian conjugate of the initial Hamiltonian.

 It is our aim to study
spontaneous breaking of the $\mathcal{PT}$ symmetry in an exactly solvable model of dimerized spin chain. The non-Hermitian term introduced here
is a staggered magnetic field affecting the whole chain. As we will show, in order to have a region of parameters with a full real spectrum,
the nearest-neighbor spin-spin interaction needs to be staggered as well. The search for the physical bases of such non-Hermitian system
will be carried out without invoking the ``equivalent Hermitian counterpart'' as described before. Following a more physical approach,
we will study the critical behavior of the system, which can experience a (anti-) ferromagnetic quantum phase transition driven by a change
in the strength of a transverse, real, magnetic field. Then, we will look for ordinary Hermitian systems displaying the same critical behavior.
As we will see, the same phase diagram can be obtained considering isotropic XX chains, while in the presence of anisotropy the mapping is possible 
with a Hermitian chain with non-linear dependence on the intensity of the transverse magnetic field.

The action of time reversal and parity in a discrete system is summarized as follows.
While the time reversal operation $\mathcal{T}$ is such that $\mathcal{T}i\mathcal{T}=-i$, the effect
of the parity  on a system of $N$ spins is such that $\mathcal{P}\sigma _{l}^{\alpha }\mathcal{P}=\sigma _{N+1-l}^{\alpha }$,
being $\sigma _{l}^{\alpha }$ the $\alpha {\rm th}$ Pauli matrix ($\alpha=x,y,z $) acting on the $N {\rm th}$ spin.

We start our discussion by considering a nearest-neighbor, Hermitian, dimerized chain of
an even number $N$ of spin 1/2:
\begin{widetext}
\begin{equation}
H_0=\sum_{l=1}^{N/2}\sum_{i=1}^{2}( \frac{J_{i}+\gamma _{i}}{2}%
\sigma _{2l-2+i}^{x}\sigma _{2l-1+i}^{x}+\frac{J_{i}-\gamma _{i}}{2}\sigma
_{2l-2+i}^{y}\sigma _{2l-1+i}^{y}) -\frac{h}{2}\sum_{l=1}^{N}\sigma _{l}^{z},
\label{hamilton}
\end{equation}
\end{widetext}
with  $\sigma _{N+1}^{\alpha }=\sigma _{1}^{\alpha }$. This model has been recently employed to study the possibility of
 observing a completely disentangled ground state.\cite{prb,canosa2010} The total Hamiltonian
 we are going to study is given by $H=H_0+V$, where $V$ is the energy due to an imaginary staggered transverse magnetic field and can be written as
\begin{equation}
 V=i\frac{\eta}{2}\sum_{l=1}^{N}(-1)^{l} \sigma _{l}^{z},
\end{equation} 
being $\eta$  a real number which measures the deviation of $H$ from Hermiticity. Without loss of generality,  $h$ and $\eta$ will be assumed to be positive.
It is clear that $V$ is  $\mathcal{PT}$ symmetric. Then, one should expect to observe such a symmetry in all 
eigenstates of $H$. As we will see, if $\eta$ is strong enough to make complex 
some of the eigenvalues of $H$, this symmetry is spontaneously broken.

The total Hamiltonian $H$ is exactly solvable, as shown in Ref.~\onlinecite{perk}. Neglecting boundary
 conditions (inessential in the thermodynamic limit), the solution can be obtained through a sequence
 of operations. The first step is the introduction of the Jordan-Wigner transformation,
defined through $\sigma _{l}^{z}=1-2c_{l}^{\dagger }c_{l}$, $\sigma
_{l}^{+}=\prod_{j<l}\left( 1-2c_{l}^{\dagger }c_{l}\right) c_{l}$, and $%
\sigma _{l}^{-}=\prod_{j<l}\left( 1-2c_{l}^{\dagger }c_{l}\right)
c_{l}^{\dagger }$, which  maps spins into spinless fermions.\cite{lieb} Then, the  chain can be divided
 into two sublattices ($c_{2l-1}=a_{l}$ and $c_{2l}=b_{l}$), and two separate Fourier transforms ($a_{l}=\left( N/2\right)
^{-1/2}\sum_{k}a_{k}\exp [-i\frac{4\pi kl}{N}]$ and $b_{l}=\left( N/2\right)
^{-1/2}\sum_{k}b_{k}\exp [-i\frac{4\pi kl}{N}]$) can be performed. Finally, a generalized Bogoliubov
 transformation, mixing $a_{k},a_{-k}^{\dagger },b_{k},b_{-k}^{\dagger }$, allows one to write 
\begin{equation}
 H=\sum_{0<k<\pi}\sum_{\alpha=+,-}  \Lambda_\alpha(k)\left(\xi^\dag_{k,\alpha}\xi_{k,\alpha}-\frac{1}{2}\right) ,
\end{equation} 
where $\xi^\dag_{k,\alpha,\pm}$ ($\xi_{k,\alpha,\pm}$) are fermionic creation (annihilation) operators,
and 
\begin{equation}
 \Lambda_\pm(k)=\sqrt{\lambda_k+\mu_k \pm 2\sqrt{\lambda_k \mu_k-\nu_k}},
\end{equation} 
with
\begin{eqnarray}
\lambda_k&=&h^2+\gamma_1^2+\gamma_2^2-2\gamma_1\gamma_2\cos 2k,\\
\mu_k&=&J_1^2+J_2^2+2J_1J_2\cos 2k-\eta^2,\\
\nu_k&=&(J_1\gamma_2+J_2\gamma_1)^2\sin 2k.
\end{eqnarray}
As usual in dimer-type chains, the spectrum consists of two branches separated by an energy gap. $\Lambda_+(k)$
is usually referred as optical branch, while $\Lambda_-(k)$ is known as acoustic branch.
A manifestation of the non Hermitian nature of $H$ is represented by the absence 
of orthogonality between the quasi-particle modes $\xi_{k,+}$ and $\xi_{k,-}$. A spectral decomposition in terms of orthogonal modes
would be in fact a signature of Hermiticity of $H$.

In the case of finite-size chains, we should take into account that $k$ can run on
integers or half-integers multiples of $\pi/N$, according to the parity of the number
of excitations of the eigenstate we are considering.\cite{katsura} This characteristic can be
exploited to predict the existence of a quantum phase transition in the thermodynamic limit.\cite{prb}

\begin{figure}[ht]
\includegraphics[width=7cm]{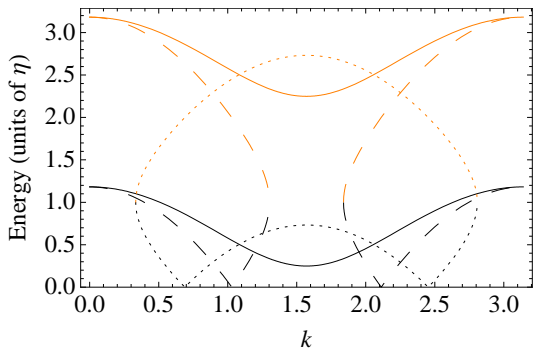}
\caption{(Color online) Band spectrum (in units of $\eta$) of the isotropic spin chain ($\gamma_1=\gamma_2=0$). Black (dark gray)
lines correspond to $\Lambda_-(k)$, while orange (light gray) lines correspond to $\Lambda_+(k)$. 
The Hamiltonian parameters are fixed as follows: (i) (solid lines) $J_1=2$, $J_2=0.4$, and $h=1$;
(ii) (dashed lines): $J_1=1.6$, $J_2=0.8$, and $h=1$; (iii) (dotted lines) $J_1=1.4$, $J_2=-0.6$, and $h=1$.
According to Eq.~(\ref{etacr}), the spectrum is fully real only in the case (i), while in the case (ii)
we have $\eta>|J_1-J_2|$ and in (iii) we have $\eta>|J_1+J_2|$. }
\label{figiso}
\end{figure}

The excitation energies take a simple expression once the anisotropy parameters $\gamma_1$ and $\gamma_2$ are set to zero.
Under these hypotheses, $\Lambda_\pm(k)$ reduce to
\begin{equation}
 \Lambda_\pm(k)=h\pm\sqrt{\mu_k}. \label{lam}
\end{equation}
Given the structure of Eq.~(\ref{lam}), it is easy to find the necessary and 
sufficient conditions under which the spectrum of $H$ is real for any $k$. 
If $J_1 J_2<0$, the function $\mu_k$ reaches its minimum 
for $k=0$ and it is negative if $\eta>|J_1+J_2|$, while when $J_1 J_2>0$ 
the minimum corresponds to $k=\pi/2$ and negative values appear whenever $\eta>|J_1-J_2|$. 
Then, we can state that all the eigenvalues of $H$ are real numbers if and 
only if $\eta<\eta_c$, where the critical value is
\begin{equation}
 \eta_c=\min\{|J_1+J_2|,|J_1-J_2|\}.\label{etacr}
\end{equation} 
Beyond this value, the breaking of the ${\cal PT}$ symmetry is observed independently 
on the size of the system. From Eq.~(\ref{etacr}) we immediately deduce that, without 
hopping dimerization ($J_1=J_2$), the ${\cal PT}$ symmetry would result unavoidably broken
and it would be impossible to obtain a full real spectrum.
In Fig. \ref{figiso}, we plot the energies of the isotropic chain for three 
different choices of parameters: in one of them the spectrum is real, while in the 
other two cases we observe the appearance of forbidden regions for the momentum $k$.
These regions build up around $0$ or $\pi/2$ depending on which of the quantities
among $|J_1+J_2|$ and $|J_1-J_2|$ is lower than $\eta$. Note that, in 
correspondence with the beginning of the ${\cal PT}$ symmetry breaking, the two branches $\Lambda_\pm$ touch.

\begin{figure}[ht]
\includegraphics[width=7cm]{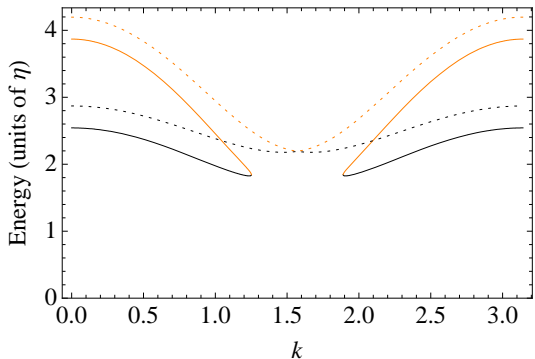}
\caption{(Color online) Band spectrum (in units of $\eta$) of the anisotropic chain. Black (dark gray)
lines correspond to $\Lambda_-(k)$, while orange (light gray) lines correspond to $\Lambda_+(k)$.
We show how the value of $h$ determines the breaking of the ${\cal PT}$ symmetry. We chose 
$J_1=1.1$, $J_2=0.1$, $\gamma_1=2.4$, and $\gamma_2=-0.8$. As for the homogeneous transverse
field, we fixed $h=0.2$ for the solid line and $h=1.5$ for the dotted line. The 
${\cal PT}$ symmetry breaking can be induced by changing the field.}
\label{figan}
\end{figure}

In the presence of anisotropic exchange ($\gamma_1,\gamma_2\neq0$), the solution for the critical value of $\eta$ is more involved.
 It is possible to state that the condition (\ref{etacr}) is still necessary but not sufficient.
The bound for the existence of a full real spectrum relies on the positivity of $\lambda_k \mu_k-\nu_k$.
 Imaginary values of $\Lambda_\pm(k)$ could in fact depend on (i) $\lambda_k \mu_k-\nu_k<0$ or on (ii) 
$\lambda_k+ \mu_k<0$. But, (ii) implies (i), and (i) does not imply (ii). The condition (i) 
can be in fact fulfilled also for positive values of $\mu_k$, being  $\lambda_k$ 
and $\nu_k$ positive by construction, while, for the condition (ii), $ \mu_k$ needs to be 
strongly negative. In Fig. \ref{figan}, we plot the excitation spectrum of $H$ 
for different values of the parameters. 
Now, also the value of the field $h$ is crucial to determine the symmetry breaking.
The disappearing of the curves coincides with the presence of imaginary roots.

In order to gain some information about the meaning of non-Hermitian Hamiltonians,
people started to study the properties of the equivalent Hermitian counterpart.\cite{mosta3,jones}
However, understanding the connection between this mathematical tool and the 
physics outlined by the model is still a demanding point. We then follow a different way,
trying to analyze the critical properties of $H$, for the region of parameters
when its spectrum is real, in comparison with those of a Hermitian systems 
spanning the same Hilbert space. 

Beyond the $\mathcal{PT}$ symmetry, $H$ is known to be invariant under the $\mathbbm{Z}_2$
group of the rotations by $\pi$ about the $z$ axis, since it commutes with the operator 
$S=\otimes_{l=1}^N \sigma_l^z$. Due to the discrete character of this symmetry, the eigenstates of $H$ are 
classified depending on the corresponding eigenvalue of $S$, which can be equal to $1$ or $-1$.
A quantum phase transition (i.e., a transition driven by the variation of an internal 
parameter of the Hamiltonian which manifests itself in a non-analytical change in some 
properties of the ground state)\cite{sachdev} takes place whenever the transverse field $h$
reaches a value such that, in the thermodynamic limit, the ground-state energies 
of the two parity sectors become degenerate. The signature of the breaking of 
the symmetry $S$ is given by a non-vanishing value of the (anti-) ferromagnetic order parameter.
A simple way to find the critical region of such models is based on the count of 
the number of excitations of each of the two symmetry sectors.\cite{katsura,prb}
Two critical fields appear that allow for negative excitations energies 
for some special $k^*$. To obtain them, we ask $\nu_{k^*}$ to vanish. Then, $k^*=0,\pi/2$. In these cases,
\begin{eqnarray}
 \Lambda_-(0)&=&\sqrt{h^2+(\gamma_1-\gamma_2)^2}-\sqrt{(J_1+J_2)^2-\eta^2},\\
 \Lambda_-(\pi/2)&=&\sqrt{h^2+(\gamma_1+\gamma_2)^2}-\sqrt{(J_1-J_2)^2-\eta^2},
\end{eqnarray} 
and the corresponding critical fields are
\begin{eqnarray}
h_C^{(1)}&=&\sqrt{(J_1+J_2)^2-\eta^2-(\gamma_1-\gamma_2)^2},\\
h_C^{(2)}&=&\sqrt{(J_1-J_2)^2-\eta^2-(\gamma_1+\gamma_2)^2}.
\end{eqnarray}
Let us assume, without loss of generality, $h_C^{(1)}<h_C^{(2)}$. 
If both of them are real, a symmetry breaking mechanism allows for 
(anti-) ferromagnetism for any field within the region $h_C^{(1)}<h<h_C^{(2)}$.
In the case of isotropic chain ($\gamma_1=\gamma_2=0$), the model belongs to a different class
of universality and a symmetry breaking mechanism takes place without the 
existence of an order parameter.

Our investigation about the relationship between our model
and the conventional Hermitian quantum mechanics begins with the isotropic model. Let us 
assume, for the sake of clarity, $J_1,J_2>0$ and $J_1>J_2$.
According to Eq.~(\ref{lam}), it is possible to define a Hermitian model, whose Hamiltonian
 will be called $H_H$, with the whole spectrum of $H$, by introducing two new exchange
 parameters $J_1^\prime$ and $J_2^\prime$ such that 
\begin{equation}
 \mu_k=J_1^{\prime2}+J_2^{\prime2}+2J_1^{\prime}J_2^{\prime}\cos 2k.\label{mueq}
\end{equation} 
 This can be achieved through $J_1^\prime=a J_1$ and $J_2^\prime=J_2/a$, where $a$ is a
renormalization parameter which diminished the difference with the two hopping amplitudes
 and then the dimer character of the chain. There exist two positive values of $a$ ($a_1$ and $a_2$) that
allow to satisfy Eq.~(\ref{mueq}). They are related by  $J_1^\prime(a_1)=J_2^\prime(a_2)$ and
$J_1^\prime(a_2)=J_2^\prime(a_1)$. Depending on the choice between $a_1$ and $a_2$, two Hermitian models
can be built that are connected with each other by a translation of one step along the chain. Furthermore,
Eq.~(\ref{mueq}) admits also $-a_1$ and $-a_2$ as possible solutions. These roots would map ferromagnets 
in anti-ferromagnets and vice-versa.

Performing this kind of analysis, the condition (\ref{etacr})
 has a simple physical interpretation, since for $\eta=\eta_c$ we find $J_1^\prime=J_2^\prime$ 
and $H_H$ represents a perfect translationally invariant chain. 
 Then, if $\eta>\eta_c$, it is not possible to identify a suitable $H_H$, given that $\eta_c$ corresponds to the maximum amount 
of renormalization we can induce in the system. In spite of the renormalization of $J_1$ and $J_2$,
the critical region of the $H_H$, bounded by $J_1^\prime-J_2^\prime$ and $J_1^\prime+J_2^\prime$,
has the same width of that of $H$.

By releasing the hypothesis of isotropy, i.e., by considering $\gamma_1,\gamma_2\neq0$, the renormalization
procedure described before is not enough to find a Hermitian model $H_H$ sharing with $H$ the same spectrum,
due to the dependence on $J$ of $\nu_k$. We are however naturally led to modify also the 
anisotropy parameters $\gamma$ to preserve $\nu_k$ from the renormalization, by defining
$\gamma_1^{\prime}=\gamma_1/a$ and $\gamma_2^{\prime}=a\gamma_2$. Due to this transformation,
$\lambda_k$ is also modified. Then, a mapping can be realized from $H(h)$ to $H_H(h^{\prime})$ 
with $h^{\prime}$ such that $h^2+\gamma_1^2+\gamma_2^2=h^{\prime 2}+\gamma_1^{\prime2}+\gamma_2^{\prime2}$.
Thus, we have found a Hermitian operator with the same spectrum of $H$ subject to 
a transverse field of non-linear strength.

To resume, we have introduced a non-Hermitian, ${\cal PT}$-symmetric, dimerized spin chain,
and found the conditions under which this model admits a fully real energy spectrum. 
In an effort to understand the physical meaning of this non Hermitian system, we have analyzed
its critical properties studying the (anti-) ferromagnetic quantum phase transition and looked for
Hermitian models exhibiting the same phase diagram. Considering an isotropic chain, the Hermitian
counterpart can be obtained scaling the hopping parameters. 
In the absence of dimerization, the Hermitian counterpart cannot be found independently of the strength of the non-Hermitian term. 
Considering an anisotropic chain, 
also the anisotropy parameters need to be renormalized. As a consequence, the Hermitian Hamiltonian
has a non-linear dependence on the strength of the transverse magnetic field.

\acknowledgments This work was partially funded by CoQuSys under project No. 200450E566. The author is supported by the Spanish Ministry of Science and Innovation 
through the program Juan de la Cierva.


\begin{thebibliography}{10}


\bibitem{bender} C. M. Bender and S. Boettcher, Phys. Rev. Lett. {\bf 80}, 5243 (1998).

\bibitem{mosta1} A. Mostafazadeh, J. Phys. A {\bf 36}, 7081 (2003). 

\bibitem{mosta2} A. Mostafazadeh, J. Phys. A {\bf 37}, 11645 (2004). 

\bibitem{benderrev} C. M. Bender, Rep. Prog. Phys. {\bf 70} 947 (2007).

\bibitem{jin} L. Jin and Z. Song, Phys. Rev. A {\bf 80}, 052107 (2009).

\bibitem{jin2}  L. Jin and Z. Song, Phys. Rev. A {\bf 81}, 032109 (2010).

\bibitem{korff} C. Korff,  J. Phys. A {\bf 41}, 295206 (2008).

\bibitem{kw} C. Korff and R. Weston, J. Phys. A {\bf 40}, 8845 (2007).

\bibitem{castro} O. A. Castro-Alvaredo and A. Fring, J. Phys. A {\bf42}, 465211 (2009).

\bibitem{mosta3} A. Mostafazadeh, J. Math. Phys. {\bf 43}, 205 (2002). 

\bibitem{prb} G. L. Giorgi, Phys. Rev. B {\bf 79}, 060405(R) (2009).

\bibitem{canosa2010} N. Canosa, R. Rossignoli, and J. M. Matera, Phys. Rev. B {\bf 81}, 054415 (2010).

\bibitem{perk}  J. H. H. Perk, H. W. Capel, M. J. Zuilhof and Th. J.
Siskens, Physica A {\bf 81}, 319 (1975).

\bibitem{lieb}  E. Lieb, T. Schultz and D. Mattis, Ann. Phys. (N.Y.) {\bf 16}, 407 (1961).

\bibitem{jones} H. F. Jones, J. Phys. A {\bf 38}, 1741 (2005). 

\bibitem{sachdev} S. Sachdev, {\it Quantum Phase Transitions} (Cambridge University Press, Cambridge, England, 1999).

\bibitem{katsura}  S. Katsura, Phys. Rev. {\bf 127}, 1508 (1962).

\end{thebibliography}
\end{document}